\documentclass[twocolumn,10pt,letterpaper]{revtex4}
\usepackage{psfrag}
\usepackage{amssymb}
\usepackage{indentfirst}
\usepackage[dvips]{graphicx}
 \usepackage{setspace}

\begin{document}

\title{Near-field induction heating of metallic nanoparticles due to infrared magnetic dipole contribution}
\author{Pierre-Olivier Chapuis}
\author{Marine Laroche}
\author{Sebastian Volz}
\author{Jean-Jacques Greffet}
\affiliation{Laboratoire d'Energ\'etique Mol\'eculaire et Macroscopique, Combustion\\CNRS UPR 288, Ecole Centrale Paris\\Grande Voie des Vignes, F-92295 Ch\^atenay-Malabry cedex, France\\
E-mail: olivier.chapuis@em2c.ecp.fr}
\linespread{1}
\begin{abstract}
We revisit the electromagnetic heat transfer between a metallic nanoparticle and a metallic semi-infinite substrate, commonly studied using the electric dipole approximation. For infrared and microwave frequencies, we find that the magnetic polarizability of the particle is larger than the electric one. We also find that the local density of states in the near field is dominated by the magnetic contribution. As a consequence, the power absorbed by the particle in the near field is due to dissipation by fluctuating eddy currents. These results show that a number of near-field effects involving metallic particles should be affected by the fluctuating magnetic fields. 
\end{abstract}

\maketitle

\section{Introduction}

{A lot of attention has recently been devoted to the interaction between nano-objects like atoms, nanoparticles or AFM tips, and surfaces, which is mediated by fluctuating thermal fields. A great variety of phenomena like Casimir-Polder forces \cite{Casimir_Polder,Casimir_Polder2,Casimir_Polder3,Casimir_Polder4}, friction forces \cite{Friction,Friction2,Friction3,Friction5,Friction4} or near-field heat transfer \cite{Carsten_Piege,RHT_Plan,RHT_Plan2,RHT_Plan3,RHT_Plan4,RHT_Plan5,RHT_Plan6, Plan_Dipole,Plan_Dipole2, RHT_Plan_Spectrum,RHT_Plan_Spectrum2,Dipole,Dipole2,Dipole3,Dipole4,Mulet2001,Pendry2} is governed by the associated stochastic thermal currents. A common assumption is that the electric dipole approximation can be used to model the nano-object \cite{Dipole,Dipole2,Dipole3,Dipole4,Pendry2,Plan_Dipole,Plan_Dipole2,Mulet2001}. Here, we revisit the heat transfer between a surface and a metallic nanoparticle. We find that the leading mechanism is near-field induction heating, due to Joule dissipation of eddy currents in the particle. The large currents are produced by time-dependent infrared magnetic fields that dominate the energy density near a metallic surface. We find a different distance dependence of the flux as compared with the case of polar materials. Our work may find applications on local heating for data storage \cite{Hamann}, and lithography \cite{Pendry2,CostasAPL}.

All the phenomena previously cited have to be described in the framework of fluctuational electrodynamics introduced by Rytov \cite{Rytov}. It is known that the radiative heat flux between two bodies \cite{RHT_Plan,RHT_Plan2,RHT_Plan3,RHT_Plan4,RHT_Plan5,RHT_Plan6, Plan_Dipole,Plan_Dipole2, RHT_Plan_Spectrum,RHT_Plan_Spectrum2} can be dramatically enhanced when their separation distance becomes smaller than $10 \mu$m. It was found that evanescent waves yield the leading contribution to the heat flux. Experiments have been reported demonstrating these effects \cite{Experiments,Experiments2,Kittel}. 
It has also been predicted that this heat transfer could have a very narrow energy spectrum \cite{Mulet2001,RHT_Plan_Spectrum,RHT_Plan_Spectrum2} due to surface electromagnetic waves. A possible application to design near-field energy converters has been studied \cite{Converters}.

The electric dipole moment of a sphere with radius $R$ and dielectric constant $\epsilon_{r}$ is generally assumed to give the leading contribution \cite{Dipole,Dipole2,Dipole3,Dipole4,Pendry2,Plan_Dipole,Plan_Dipole2,Mulet2001} because it varies like $(R/\lambda)^3$ whereas the next term in the Mie expansion varies as $(R/\lambda)^5$ ($\lambda$ is the wavelength in vacuum) \cite{VanDeHulst}. In this work, we will show that the interaction between the magnetic dipole and the large magnetic fields in the near field  may give the dominant contribution to the heat transfer. 

In the next section, we compare the absorption cross section of the electric and magnetic dipole moment. The third section is devoted to the analysis of the electric and magnetic energy density in the near field of a metal-vacuum interface. The final section analyses the heat transfer and discusses the physical mechanism.

\section{Absorption by a metallic nanoparticle}

Let us first compute the power absorbed by a small metallic particle. In what follows, we will use an isotropic, homogeneous and local form of the complex dielectric constant. To lowest order in $R/\lambda$ \cite{VanDeHulst}, the particle can be described by its electric dipolar moment $\vec{p}$. We define a complex polarizability $\alpha_{E}$ 
\begin{eqnarray}
\vec{p} =\alpha_{E}~\epsilon_{0}~\vec{E},
\end{eqnarray}
where $\epsilon_{0}$ is the dielectric permittivity in vacuum and $\vec{E}$ is the external electric field. Another contribution is given by the magnetic dipolar moment $\vec{m}$ characterized by its magnetic polarizability $\alpha_{H}$
\begin{eqnarray}
\vec{m} =\alpha_{H}~\vec{H}, 
\end{eqnarray}
where $\vec{H}$ is the external magnetic field. Higher multipoles can be neglected if $\vert\epsilon_{r}\vert \gg1$ and $R/\lambda \ll1$. The contributions of the electric and magnetic dipoles to the power dissipated in the particle at a positive frequency $\omega$ are given by \cite{Landau,JoulainReview} 
\begin{eqnarray}
P_{abs}^{E}(\omega)=\omega~2Im(\alpha_{E})~~\epsilon_{0} \frac{<|\vec{E}|^{2}>}{2},
\end{eqnarray}
\begin{eqnarray}
P_{abs}^{M}(\omega)=\omega~2Im(\alpha_{H})~~\mu_{0}\frac{<|\vec{H}|^{2}>}{2}
\end{eqnarray}
where $\mu_{0}$ is the magnetic permittivity in vacuum. $\alpha_{E}$ and $\alpha_{H}$ can be found in ref. \cite{Mulholland} :
 \begin{eqnarray}
\alpha_{E}=4\pi~R^{3}~\frac{\epsilon_{r}-1}{\epsilon_{r}+2},
\end{eqnarray}
\begin{eqnarray}
\alpha_{H}=\frac{2\pi}{15}~R^{3}\left(\frac{2\pi R}{\lambda}\right)^2(\epsilon_{r}-1),
\end{eqnarray}
where $\epsilon_{r}$ is the relative dielectric permittivity.

Here, we do not take into account the diamagnetism of the material. Instead, the magnetic dipole moment is due to eddy currents in the particle. The polarisabilities are calculated assuming that $R$ is much smaller than the skin depth $\delta$. A different form \cite{Landau} can be derived when dealing with particles such that $\delta \ll R \ll \lambda$. In what follows, we should keep in mind that the dipole model is a fair approximation provided that the distance $d$ between the center of the particle and a surface is much larger than $R$. Note that we have used for simplicity the extinction cross section of the elecric dipole. The exact form of the absorption cross section is discussed in ref. \cite{Carminati}. The difference for a metallic nanoparticle is negligible.

As seen from Eqs (3,4), the absorption is the product of two terms, the imaginary part of the polarizability and the local density of energy. We shall show that for metallic nanoparticles at low frequencies, both terms are larger for the magnetic contribution. Let us first analyse the role of the polarizability. It appears from Eqs (3-6) that for values of the dielectric constant on the order of unity, the electric dipole contribution to losses is much larger than the magnetic one because $R/\lambda \ll1$. Yet, for values of $\epsilon_{r}$ such that $\vert\epsilon_{r}\vert \gg1$, as it is the case for metals at low frequencies, the magnetic dipole may provide the leading contribution. The physical reason is that the magnetic fields are continuous at an interface so that they can penetrate in the material. By contrast, the electric field in a spherical particle $\vec{E}_{int}$ is related to the external electric field by $\vec{E}_{int}=[3/(\epsilon_{r}+2)]\vec{E}_{ext}$. Surface charges induced at the interface prevent the electric field to penetrate efficiently in the metallic particle. This screening effect takes place on a length scale given by the Thomas-Fermi length. It does not depend on the skin depth. 
\begin{figure}[h]
\includegraphics[width=7.5cm]{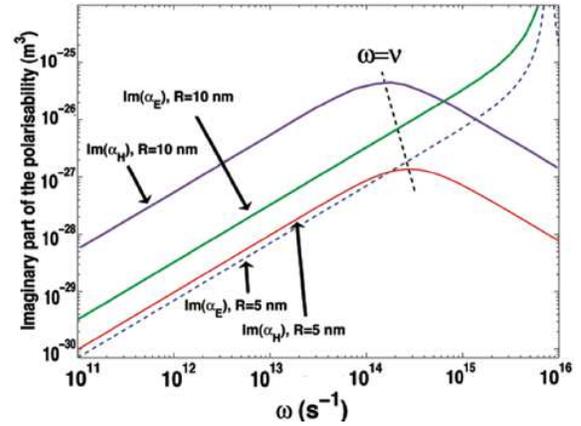}
\caption{Imaginary parts of the electric and magnetic polarisabilities of a gold sphere ($\omega_p=1.71~10^{16}$ s$^{-1}$, $\nu_{0}=4.05~10^{13}$ s$^{-1}$, $v_{F}=1.2~10^6$ms$^{-1}$ and $A=1$). For $\omega \ll \nu$, $Im({\alpha_{H})\approx}\frac{4\pi R^{5}\omega_{p}^{2}}{15c^{2}}\frac{\omega}{\nu}$ and for $\omega \ll \omega_p$, $Im({\alpha_{E})\approx}\frac{12\pi R^{3}\nu}{\omega_{p}^{2}}~\omega$ } 
\label{default}
\end{figure}

 We consider a non-magnetic metallic particle characterized by a Drude model $
\epsilon_{r}=1-\omega_{p}^{2}/(\omega^2+i~\omega \nu) $ where $\omega_{p}$ is the plasma frequency and $\nu$ is the damping coefficient. To account for the confinement effects, the bulk dielectric constant $\epsilon_{r}$ \cite{Kreibig} is corrected by modifying the damping constant $\nu = \nu_{0}+A~v_{F}/R$, where $\nu_{0}$ is the bulk damping coefficient, $v_{F}$ the Fermi velocity and $A$ a sample-dependent coefficient. Figure 1 shows $Im(\alpha_{E})$ and $Im(\alpha_{H})$ as a function of circular frequency for two gold spheres with radii $R=5$ nm and $R=10$ nm. It is seen that the electric polarizability is larger  than the magnetic polarizability at optical frequencies.  As explained before, this is no longer the case at low frequencies (typically smaller than $\nu$), where $Im(\alpha_{H})$ is larger than $Im(\alpha_{E})$.

\section{Local density of energy near a metallic surface}

To derive the energy absorbed by a particle in the vicinity of an interface, we need to consider the local densities of energy $\epsilon_{0} \frac{<|\vec{E}|^{2}>}{2}$ and $\mu_{0} \frac{<|\vec{H}|^{2}>}{2}$. In a vacuum, both contributions are equal. The energy per unit volume $U(z,\omega)$ at a distance $z$ from the interface increases dramatically in the near field due to the presence of evanescent waves as discussed in ref. \cite{Joulain2003,JoulainReview}. $U(z,\omega)$ is the product of the local density of states (LDOS) $\rho(z,\omega)$ by the mean energy of a mode  $\Theta (\omega,T)=\hbar \omega/[\exp(\hbar \omega/k_{B} T)-1]$, 
where $2\pi \hbar$ is Planck constant, $k_{B}$ is Boltzmann constant and $T$ the temperature of the substrate. The final expression for the evanescent part of the LDOS \cite{Joulain2003,JoulainReview} is the sum of the four following contributions :
\begin{eqnarray}
\rho^{E}_{s}(z,\omega)&=&\rho_{v}\int_{\omega/c}^{+\infty}\frac{dK}{2|\gamma_{0}|}\frac{cK}{\omega}Im(r_{s})e^{-2\gamma_{0}"z}  \\
\rho^{M}_{s}(z,\omega)&=&\rho_{v}\int_{\omega/c}^{+\infty}\frac{dK}{2|\gamma_{0}|}\frac{cK}{\omega}f(K,\omega)Im(r_{s})~e^{-2\gamma_{0}"z}
 \\
 \rho^{E}_{p}(z,\omega)&=&\rho_{v}\int_{\omega/c}^{+\infty}\frac{dK}{2|\gamma_{0}|}\frac{cK}{\omega}f(K,\omega)Im(r_{p})~e^{-2\gamma_{0}"z}
 \\
\rho^{M}_{p}(z,\omega)&=&\rho_{v}\int_{\omega/c}^{+\infty}\frac{dK}{2|\gamma_{0}|}\frac{cK}{\omega}Im(r_{p})e^{-2\gamma_{0}"z}
\end{eqnarray}
where the subscripts $^{E}$ and $^{M}$ denote the electric and magnetic evanescent components, $c$ is the light velocity in vacuum, $\rho_{v}(\omega)=\omega^{2}/\pi^{2}c^{3}$ is the vacuum density of states, $f(K,\omega)=2(\frac{cK}{\omega})^{2}-1$, $r_{s}=\frac{\gamma_1-\gamma_0}{\gamma_1+\gamma_0}$ and $r_{p}=\frac{\epsilon_1\gamma_0-\epsilon_2 \gamma_0}{\epsilon_1 \gamma_0-\epsilon_0 \gamma_1}$ the Fresnel TE and TM reflection factors, and the complex number $\gamma_{i}=\gamma_{i}'+i~\gamma_{i}"$ is defined as the perpendicular part of the wave vector at a frequency $\omega$ : $K^{2}+\gamma_{0}^{2} = \epsilon_{i} \frac{\omega^2}{c^2}$ where $i=0$ in vacuum ($\epsilon_{0}=1$) and $i=1$ in the metal. We have neglected non-local effects as we consider distances larger than the Thomas-Fermi screening length.

\begin{figure}[h]
\includegraphics[width=7.5cm]{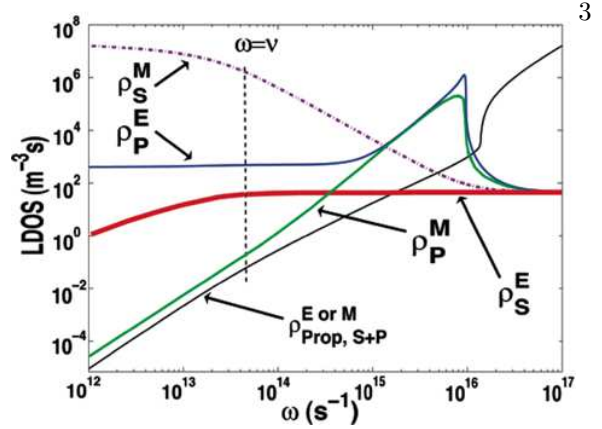}  
\caption{Contributions of the evanescent waves to the local density of states (LDOS) at $d=30$~nm of a gold-vacuum plane interface when using a bulk Drude dielectric constant. The magnetic and electric propagating contributions are also plotted (they are equal).}
\label{default}
\end{figure}

On Figure 2, we plot the LDOS versus the frequency $\omega$ for $d=30$~nm. The first conclusion is that the contribution due to the evanescent waves dominates. The second conclusion is that the s-polarized magnetic contribution is dominant for frequencies below $\omega_M=2.4~10^{14}$s$^{-1}$ which are relevant for heat transfer at $300$~K. Indeed, in the expression of the energy density $U(z,\omega)$, $\Theta (\omega,T)$ acts as a temperature-dependent frequency filter. At a given temperature, we define a cut-off frequency $\omega_M$ by $\int_0^{\omega_M}\Theta(\omega,T)d\omega/\int_0^{\infty}\Theta(\omega,T)d\omega =99/100$. Frequencies much higher than $\omega_M$ are not relevant for heat transfer. We note that the p-polarized contribution associated with the surface plasmon polariton dominates at optical frequencies but does not contribute significantly in the infrared. 

The dominant contribution of magnetic energy can be understood by considering the analytical expressions of $\rho^{M}_{s}$ (Eq. 8) and  $\rho^{E}_{p}$ (Eq. 9). Both expressions are exactly symmetric, involving the same factor $f(K,\omega)$, and the imaginary part of the reflection factor, respectively $Im(r_s)$ and $Im(r_p)$. 
The physical origin of the factor $f(K,\omega)$ lies in a fundamental difference of structure between propagating and evanescent waves. It is interesting to see why the magnetic energy dominates the electric energy. Indeed, in a vacuum,  the electric and the magnetic energy are equal. This is no longer true for an s-polarized evanescent wave close to an interface.

Let us denote the wave vector as follows
\begin{eqnarray}
\vec{k_0}=K~\vec{e_x}-\gamma~\vec{e_z}
\label{eq.23}
\end{eqnarray}
where $K$ and $\gamma$ are the interface parallel and perpendicular wave vectors and $\vec{e_x}$ and $\vec{e_z}$ unit vectors. The Helmholtz equation in a vacuum yields $K^2+\gamma^2=\omega^2/c^2$.  For an s-polarized field, the electric field is given by
\begin{eqnarray}
\vec{E}=(0,E,0),
\label{eq.24}
\end{eqnarray}
and the magnetic field follows from  the Maxwell-Faraday equation in vacuum ($\vec{curl}~\vec{E}=-\frac{\partial}{\partial t} \vec{B}$)
\begin{eqnarray}
\vec{B}=\frac{E}{\omega}(-\gamma,0,K).
\label{eq.25}
\end{eqnarray}
It follows that $|\vec{B}|^2=\vec{B}\vec{B}^{*}=\frac{|E|^2}{\omega^2}(|\gamma|^2+K^2)$. For a propagating wave, this yields the well-known result
\begin{eqnarray}
|\vec{B}|=c~|\vec{E}|,
\label{eq.26}
\end{eqnarray}
whereas for an evanescent wave, $\gamma$ is purely imaginary so that $|\gamma|^2=-\gamma^2$. We get:
\begin{eqnarray}
\frac{|\vec{B}|}{c~|\vec{E}|}= \sqrt{2\frac{K^2}{k_0^2}-1}.
\label{eq.27}
\end{eqnarray}
For evanescent waves, $K\gg k_0$ so we find that the magnetic energy stored in an s-polarized evanescent wave is much larger than the electric energy.
We have plotted in Figure \ref{fig:FonctionF} the function $\frac{B}{cE}=f(K, \omega)=\sqrt{\frac{|\gamma|^2+K^2}{k_0^2}}$ for different frequencies. It is seen that the density of energy, which is proportional to the density of states, is driven by the magnetic contribution.
\begin{figure}[h]
\includegraphics[width=7.5cm]{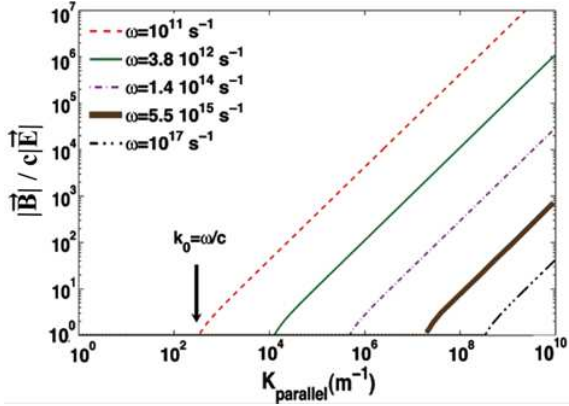}
\caption{Ratio $\frac{|\vec{B}|}{c|\vec{E}|}$ for s-polarized evanescent waves.}
\label{fig:FonctionF}
\end{figure}

Now, a similar reasoning can be done for p-polarized waves. In this case, the inverse ratio $\vert \vec{E}\vert/c\vert \vec{B}\vert$ is also equal to $\sqrt{f(K,\omega)}\simeq \sqrt{2}~K/k_0$ which shows that the electric field dominates in this case. Near field is thus always dominated by an s-polarized evanescent magnetic field and a p-polarized evanescent electric field. 
The relative weight of both contributions is then given by the values of the imaginary part of the reflection factors. Since $Im(r_s)$ is larger than $Im(r_p)$ for a metal at low frequencies, the s-polarized contribution to the LDOS dominates as seen in Figure 2. As a consequence, the energy density is dominated by its s-polarized magnetic  contribution. 

\section{Heat transfer between an interface and a nanoparticle}

We now combine the results obtained for the dependence of the polarisabilities and for the energy density to derive the power absorbed by a small particle as given by Eqs (3,4). Although heat transfer between a small particle and a substrate in the near field has only been calculated using the electric dipolar contribution \cite{Dipole,Dipole2,Dipole3,Dipole4,Pendry2,Plan_Dipole,Plan_Dipole2,Mulet2001}, the results shown in Figure 1 (large magnetic dipole moment) and in Figure 2 (large magnetic density of states) clearly indicates that the magnetic contribution must be taken into account as suggested in ref. \cite{Kittel}. Figure 4 shows the radiative power 
\begin{eqnarray}
P_{rad}=\int_{\omega=0}^{+\infty}(P_{abs}^{E}(\omega)+P_{abs}^{H}(\omega))d \omega
\end{eqnarray}
dissipated by the substrate in the small particle ($R=5~$nm). The key result observed in Figure 4 is that the heat transfer is dominated by the s-polarized magnetic contribution. The magnetic contribution can be  larger than the electric dipolar contribution by 3 orders of magnitude. The reason is that heat is dissipated essentially by eddy currents. An important result is the dependence of the heat flux with distance. The magnetic LDOS varies asymptotically as $1/z$ and the electric LDOS as $1/z^3$ \cite{JoulainReview,Joulain2003}.  For gold, these behaviors are valid at very small distances (below $20$~nm). Hence, there is no simple distance dependence for the absorbed power as seen in Figure 4. 
 
  \begin{figure}[h]
\includegraphics[width=7.5cm]{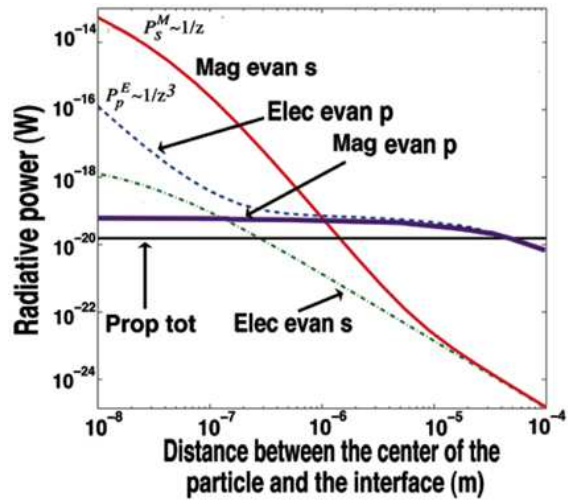}
\caption{Radiative power dissipated in the gold particle (radius $R=5$~nm) by the semi-infinite planar gold substrate at $300$~K, and the asymptotic behaviours}
\label{default}
\end{figure}

The above analysis can be summarized by the following scenario. Random currents flowing parallel to the interface can excite the s-polarized evanescent electromagnetic fields at infrared frequencies. As explained above, the associated magnetic fields take large values in the near-field. They are continuous across a vacuum-metal interface so that they penetrate efficiently in the nanoparticle and can generate large eddy currents. These currents are dissipated through Joule effect. Thus, thermal heat transfer appears to be due to near-field induction heating. 
Radiative heat transfer between two parallel metallic surfaces can also be revisited with a similar scenario \cite{Chapuis}.

 \section{Conclusion}
 
In summary, we have shown that the heat transfer between a metallic nanoparticle and a metallic surface is dominated by the magnetic contribution. Heat is mainly dissipated by fluctuating eddy currents. The widely used electric dipole approximation is valid for dielectrics but breaks down for metals. As a consequence, the $1/z^{3}$ dependence of the flux between dielectrics is not valid for metals. A number of other effects due to thermal radiation (e.g. forces, friction) between metallic bodies are expected to be driven by their magnetic contribution, even if the media are non-magnetic. We note that the heat exchanged by two metallic nanoparticles separated by a submicronic distance \cite{Domingues}  should be driven as well by the interaction between their magnetic dipoles. 

\section*{Acknowledgments}
We thank Karl Joulain and Carsten Henkel for useful discussions. We acknowledge the support of the Agence Nationale de la Recherche under contract ANR06-NANO-062-04.
{
\small
\linespread{1}

}
 \end{document}